\documentclass[prl,twocolumn,aps,showpacs,superscriptaddress,epsfig]{revtex4-1}
\usepackage{amssymb}
\usepackage{amsbsy}
\usepackage{amsmath}
\usepackage{epsfig}
\usepackage{graphicx}

\usepackage{hyperref}
\hypersetup{
    colorlinks=true,
    linkcolor=blue,
    filecolor=cyan,      
    urlcolor=magenta,
    citecolor=red,
}
\begin{document}

\title {Non-linear transport without spin-orbit coupling or warping in two-dimensional Dirac semimetals}
\author{Sai Satyam Samal}
\affiliation{National Institute of Science Education and Research, Jatni, Odisha 752050, India}
%\affiliation{Homi Bhabha National Institute, Training School Complex, Anushakti Nagar, Mumbai 400094, India }
\author{S. Nandy}
\affiliation{Department of Physics, University of Virginia, Charlottesville, VA 22904, USA}
\author{Kush Saha}
\affiliation{National Institute of Science Education and Research, Jatni, Odisha 752050, India}
\affiliation{Homi Bhabha National Institute, Training School Complex, Anushakti Nagar, Mumbai 400094, India }
%\date{\today}

\begin{abstract}
%Berry phases are known to have substantial influence on the non-linear transport in materials without certian symmetries. Specifically, the first moment of Berry curvature, namely Berry dipole gives rise to non-linear current in a wide variety of time-reversal invariant and noncentrosymmetric materials. 
It has been recently realized that the first-order moment of the Berry curvature, namely the \textit{Berry curvature dipole} (BCD) can give rise to non-linear current in a wide variety of time-reversal invariant and non-centrosymmetric materials. While the BCD in two-dimensional Dirac systems is known to be finite  only in the presence of either substantial spin-orbit coupling where low-energy Dirac quasiparticles form tilted cones or higher order warping of the Fermi surface, we argue that the low-energy Dirac quasiparticles arising from the merging of a pair of Dirac points without any tilt or warping of the Fermi surface can lead to a non-zero BCD. Remarkably, in such systems, the BCD is found to be independent of Dirac velocity as opposed to the Dirac dispersion with a tilt or warping effects. We further show that the proposed systems can naturally host {\it helicity}-dependent photocurrent due to their linear momentum-dependent Berry curvatures. Finally, we discuss an important byproduct of this work, i.e., nonlinear anomalous Nernst effect as a second-order thermal response.  
\end{abstract}

\maketitle

{\em Introduction:} The Berry phases of electronic wavefunctions can substantially modify the transport properties and give rise to a plethora of anomalous transport phenomena in the linear response regime such as anomalous Hall effect (AHE), anomalous Nernst effect (ANE), quantum charge pumping, etc~\cite{Niu_2010} in systems with broken time-reversal symmetry. Recently, it has also been proposed that the first-order moment of the Berry curvature, namely the Berry curvature dipole (BCD) can give rise to non-linear current in  a time-reversal invariant and non-centrosymmetric material  ~\cite{Fu_2015,Moore_2015,Guinea_2015}. Subsequently, the discovery~\cite{Ma2019, Kang2019} of non-linear anomalous Hall effect (NLAHE) in layered transition metal dichalcogenides (TMDCs) makes the non-linear transport as one of the prime topics of interest to both theorists and experimentalists~\cite{Guinea_2015, Fu_2015,Moore_2015, Son_BCD_2019,Zhang_BCD_2018,Low_2018,Xie_2018, Niu_2019, Nandy_2019, Nandy1_2019, Sodemann_2019, Law_2020, Evgeny_2020, Du2019, Wang2019, du2020quantum,Zhang_2018, Brink_2018,Rostami} in recent times.  
%Because of the non-linear current response, the NLAHE can transform ac electric fields into dc currents, a process known as rectification, which has great potential applications for next-generation wireless and energy-harvesting devices~\cite{Fu_2020}.

%For example, non-linear Hall effect, non-linear photo current and non-linear Nernst current are three most important transport phenomena originates from the Berry curvature.
%Specifically, there has been a great deal of interest in understanding Hall-like phenomena without any external magnetic field\cite{}. While  quantized conductivity in conventional Hall effect is obtained by integrating Berry curvature within the Fermi surface, the non-linear nature of the electric driven Hall effect is obtained by integrating the first order {\it derivative} of Berry curvature over the Fermi surface.   

Unlike the linear transport properties induced by the Berry curvature, the BCD-induced non-linear transport phenomena such as non-linear anomalous Hall effect, non-linear anomalous Nernst effect (NLANE) and non-linear anomalous thermal Hall effect (NLATHE) is found to be Fermi surface quantities~\cite{Fu_2015,Nandy1_2019,Su_2019,Nandy2_2019}. To find finite values of these non-linear phenomena, one requires either low-symmetry crystals or Dirac Hamitonian with higher order corrections. For example, it has been shown that the presence of substantial spin-orbit coupling (SOC) where low-energy Dirac quasiparticles form tilted cones is necessary to find a finite NLAHE in two-dimensional (2D) Dirac systems\cite{Fu_2015}. On the other hand, a recent study~\cite{Carmine_2019} showed that the NLAHE can survive in 2D Dirac semimetals even in the absence of complete SOC but with higher order correction to the linearly dispersive Dirac Hamiltonian, particularly with warping of the Fermi surface. 

\begin{figure}
	\includegraphics[width=0.90\linewidth]{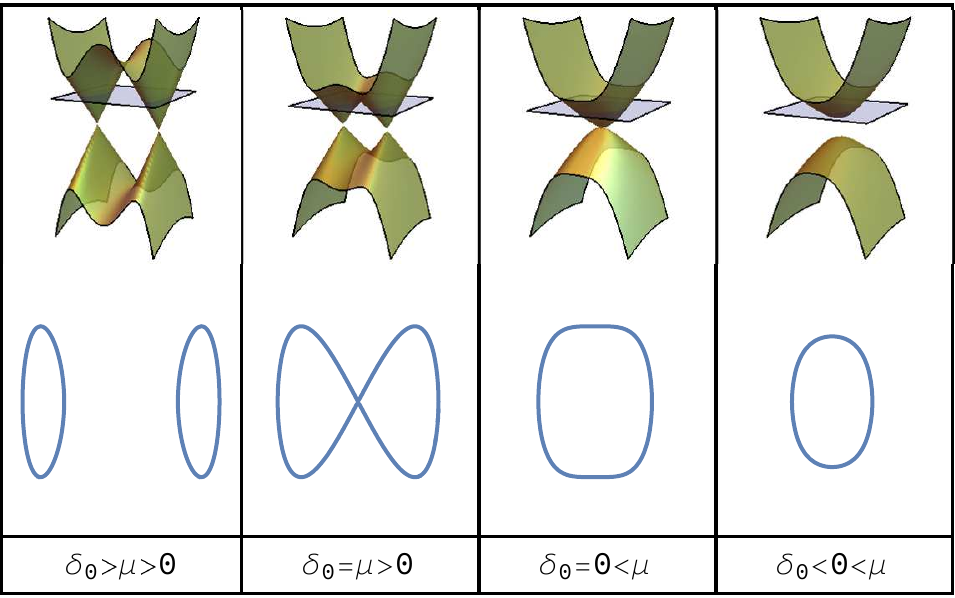}
	\caption{Evolution of energy spectrum (Eq.~\ref{eq:SemiDirac}) (top panel) and corresponding Fermi surface topologies\cite{note_1} (bottom panel) for different values of parameter $\delta_0$. For finite chemical potential $(\mu)$, the saddle point between the two Dirac nodes evolves with $\delta_0$ and crosses the chemical potential at $\delta_0=\mu$, leading to topological Lifshitz transition. Note that, for fixed $\delta_0>0$, the similar topological transition is obtained by varying $\mu$.}
	\label{fig:band}
\end{figure}

This raises an important question to address, pertinent to several recent and upcoming experiments on 2D materials: is it possible to realize a finite BCD in a simple low-energy 2D Dirac Hamiltonian {\it without} any tilting and warping effect. Remarkbly, we find that the tilting of the Dirac cone or warping of the Fermi surface is not necessary to observe a non-zero BCD in 2D DSMs. Rather, a simple low-energy Hamiltonian with a pair of Dirac nodes close to each other with a saddle point in between the nodes, or a Dirac system where two Dirac nodes merge with each other at a point turn out to be a simple platform for realizing sizeable BCD-induced non-linear Hall effect. Further, we find that the BCD is independent of Dirac velocity but predominantly depends on the effective mass parameter for a fixed band gap. Moreover, these systems can naturally give rise to {\it helicity}-dependent photocurrent due to their linear momentum-dependent Berry curvatures, as opposed to the typical gapped Dirac and semiconducting systems\cite{Moore_2015}. We finally discuss the contribution of the Berry curvature to the non-linear anomalous Nernst effect. 

Deformed graphene with uniaxial stress turns out to be an ideal platform for realizing such model Hamiltonian\cite{Goerbog_2008,Dietl_PRL_2008,Montambaux_EPJB_2009}. It has also been argued that TiO$_2$/VO$_2$ heterostructures \cite{pardo2009pickett,Montambaux_PRB_2009} under quantum confinement, (BEDT-TTF)$_2$$I_3$ organic salts under pressure \cite{katayama2006suzumura}, photonic metamaterials \cite{wu2014} can be ideal candidate materials to host these types of low energy dispersions. Recent experimental realization of such model Hamiltonian in optical lattices \cite{leticia2012esslinger} has renewed the quest for materials with tunable Dirac nodes.

%In particular, we trace out that a new class of two-dimensional anisotropic Dirac semimetals \cite{banerjee2009pickett,depplace2010montambaux,
%dietl2008montambaux}, which are known to exhibit a special phase, namely critical semi-Dirac phase characterized by electronic bands touching in a discrete set of nodes about which the bands disperse linearly in one direction and quadratically along the orthogonal direction, may give rise to sizable BCD induced non-linear anomalous transport coefficients even in the absence of SOC or warping. 

%\textcolor{red}{We furthermore show that in such systems the Berry dipole may be used as a diagonostic tool for studying toplogical liftshitz transition. To further substantiate this claim, we consider a relatively less well known models, namely Triple quarter Dirac fermions, where Fermi surface topology is rather complicated. We find that the Berry dipole exhibit a kink as a function of Fermi energy, corroborating the transition in Fermi surface topology.}

{\em Model Hamiltonian:}
The merging of a pair of Dirac nodes can be modeled by the low-energy Hamiltonian\cite{Montambaux_EPJB_2009,Adroguer_2016} 
\begin{align}
\label{eq:SemiDirac}
\mathcal{H}=\mathbf{d}(\mathbf{k})\cdot \boldsymbol \sigma,
\end{align}
where $\boldsymbol \sigma$'s are the Pauli matrices in pseudospin space and  $\mathbf{d}(\mathbf{k})=(\alpha\,k_x^2-\delta_0, v\,k_y,0)$. Here ${\bf k}=(k_x, k_y)$ is the crystal momentum,  $\alpha$=$\hbar^2/{2m}$ is the inverse of quasiparticle mass  along $x$, $v$ is the Dirac velocity along $y$, and $\delta_0$ is the parameter which drives the transition between a metallic and insulating phase. For $\delta_0>0$, two gapless Dirac nodes are found at $(\pm \sqrt{\delta_0/\alpha},0)$. At  $\delta_0=0$, the two Dirac nodes merge, leading to a special dispersion where electron disperses quadratically along the $y$ direction and linearly along the orthogonal direction. This is typically called semi-Dirac point. For $\delta_0<0$, a gapped system with trivial insulating phase is obtained. The corresponding spectrum is shown in Fig.~(\ref{fig:band}). Evidently, the saddle point between the two Dirac nodes evolves with $\delta_0$, which in turn leads to the topological Lifshitz transition for a fixed chemical potential ($\mu$) (cf. Fig.~(\ref{fig:band})). Consequently, the area of the Fermi surface $(S_{F})$ is found to vary {\it nonmonotonically} with $\delta_0$. This is expected to be reflected in the BCD since it is a Fermi surface quantity. We note that for fixed $\delta_0$, the Liftshiz transition can also be obtained with the variation of $\mu$.

Note that, Eq.~\ref{eq:SemiDirac}  obeys effective time-reversal ($\Theta=\mathcal {K}$), particle-hole ($\mathcal{P}=\sigma_y$) and chiral symmetries ($\mathcal{C}=\sigma_y \mathcal {K}$), where $\mathcal{K}$ is the complex conjugation operator. Further, we can define two effective mirror symmetries $M_x: (x,y)\rightarrow (-x,y)$ and  $M_y(x,y):(x,y)\rightarrow (x,-y)$. In momentum space, $M_x=\sigma_0$ and $M_y=\sigma_x$\cite{Kim_2017}. It is worth mentioning that Eq.~\ref{eq:SemiDirac} can be effectively obtained from a four band isotropic Dirac Hamiltonian with an anisotropy along $x$ direction\cite{fran} without any tilt.
\begin{figure}
	\includegraphics*[width=0.40\linewidth]{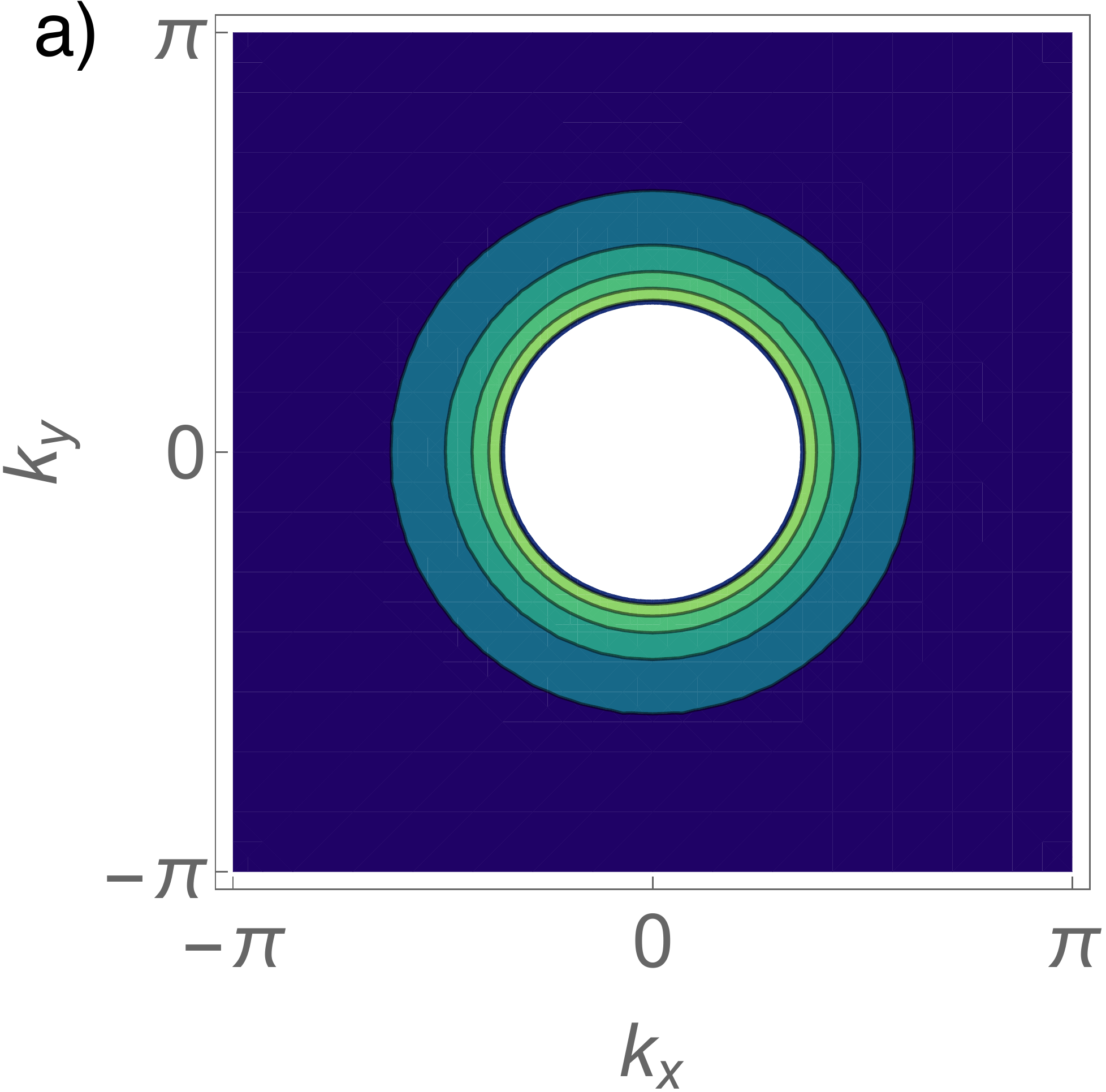}
	\includegraphics*[width=0.40\linewidth]{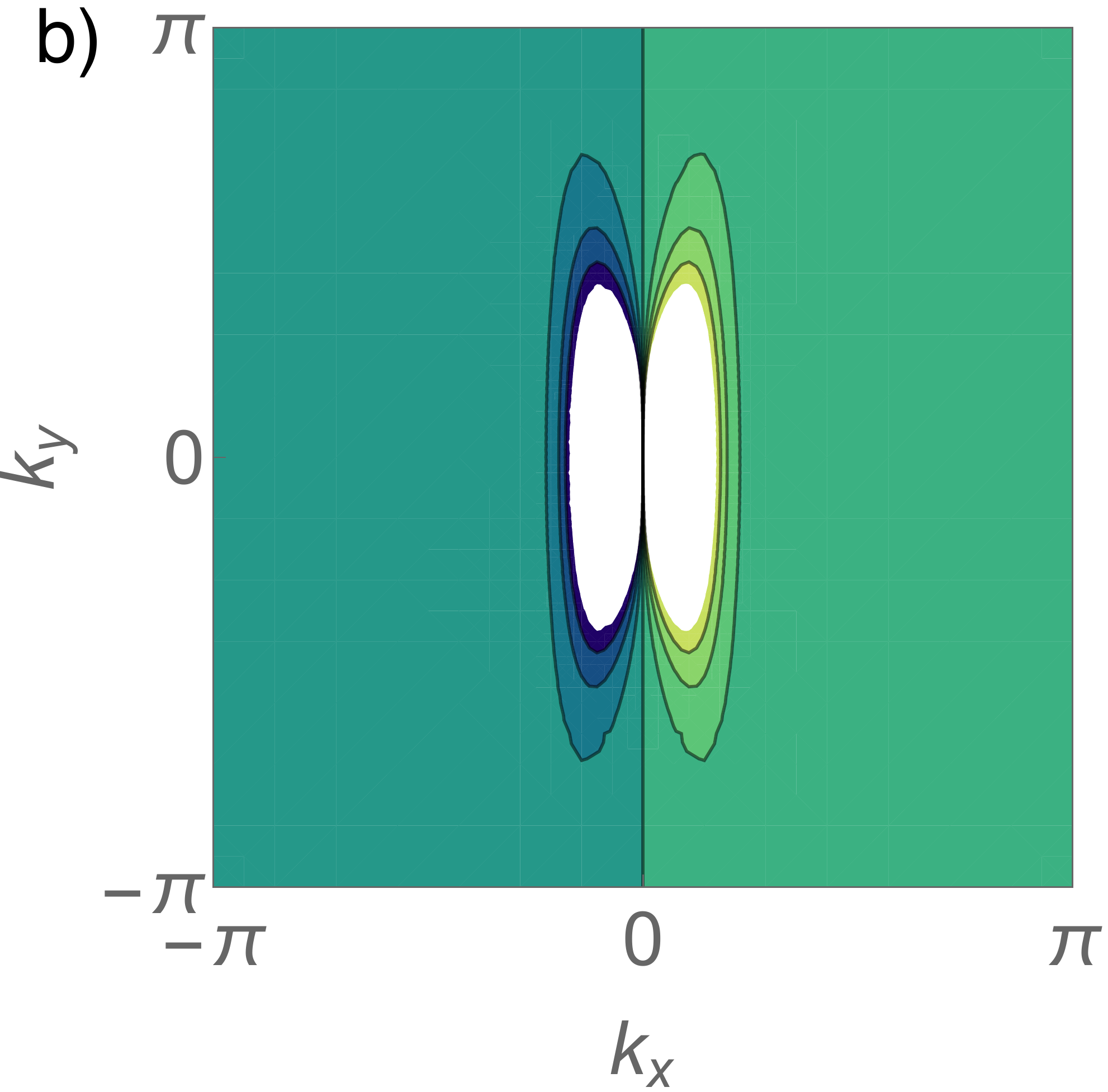}
	\includegraphics*[width=0.10\linewidth]{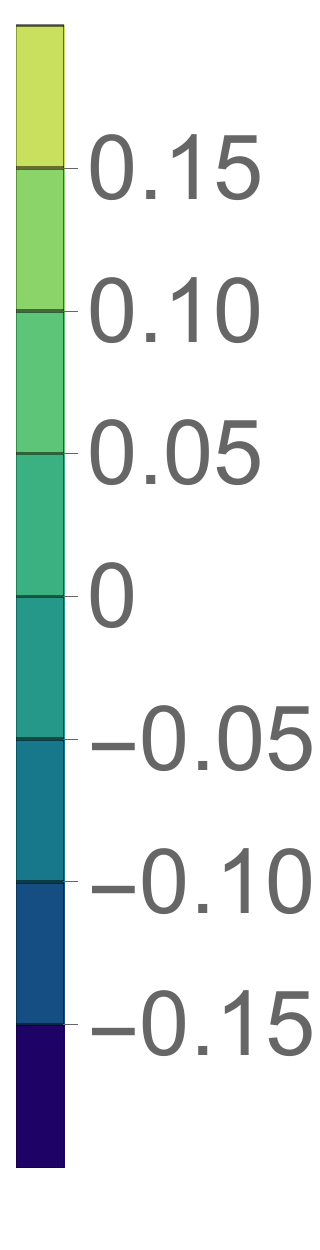}
	\includegraphics*[width=0.40\linewidth]{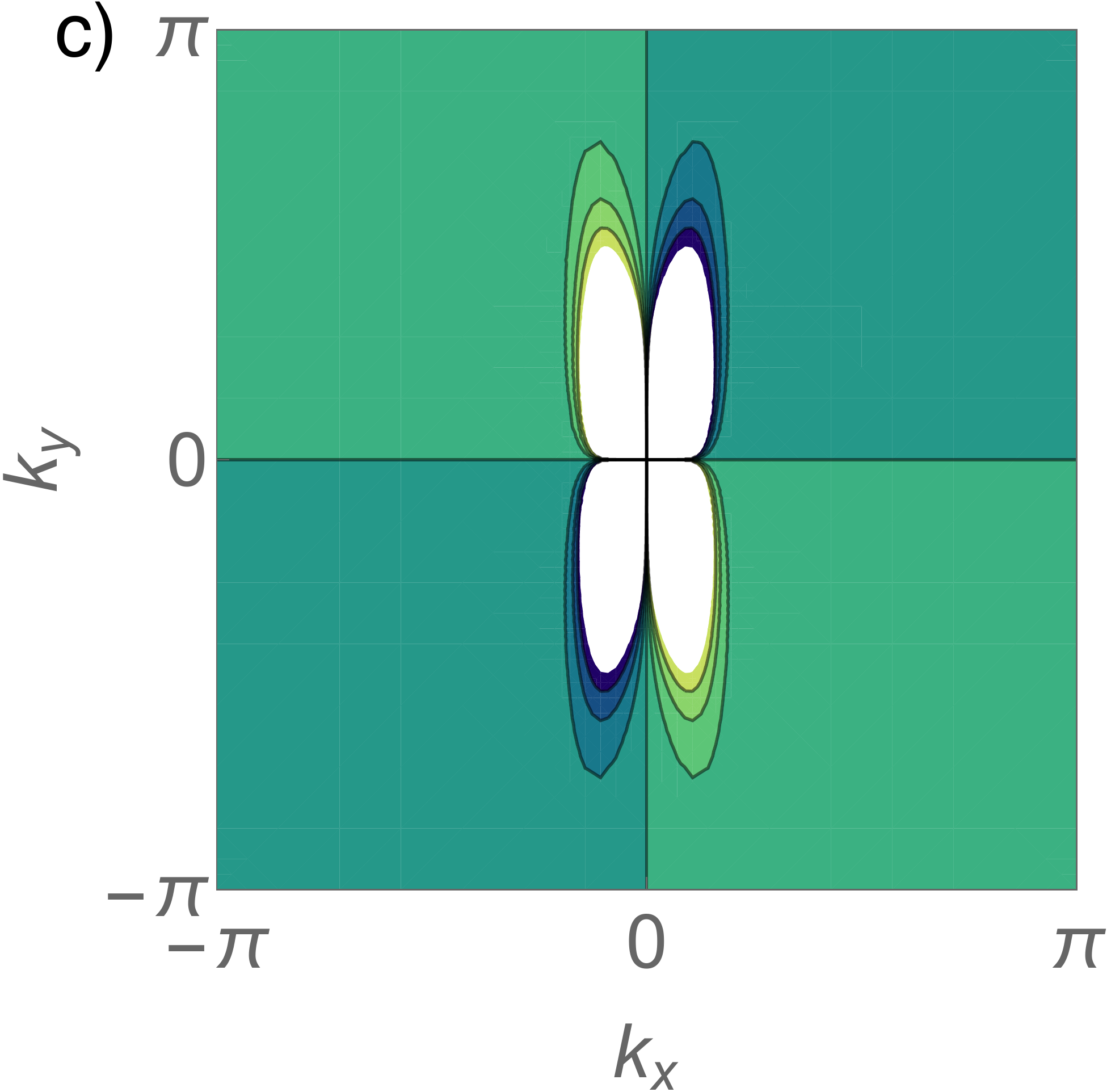}
	\includegraphics*[width=0.40\linewidth]{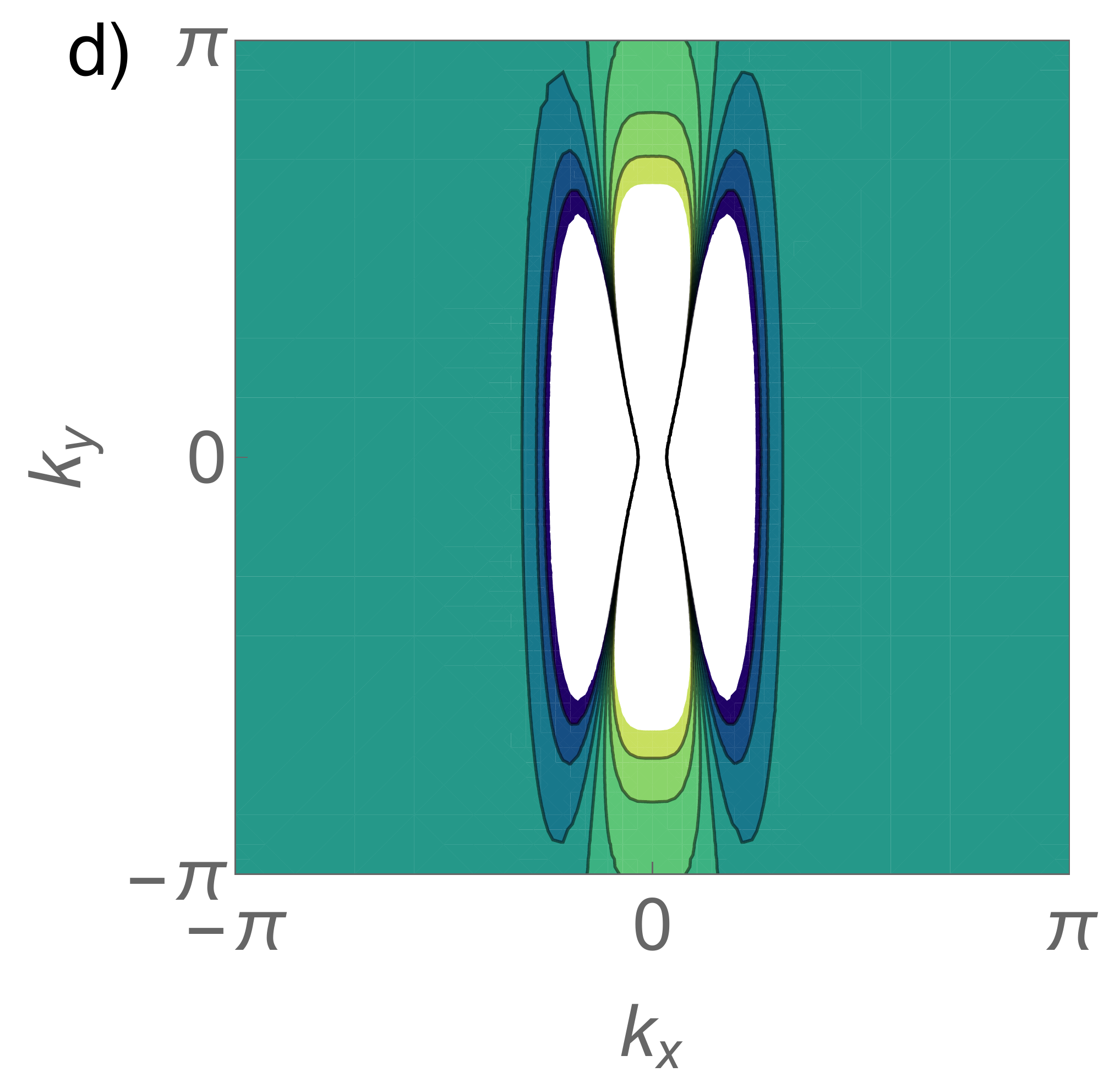}
	\includegraphics*[width=0.10\linewidth]{Legend.pdf}
	\caption{Top Panel: (a) Contour plot of Berry curvature for isotropic gapped Dirac Hamiltonian. (b) The same plot for the gapped semi-Dirac Hamiltonian discussed in the text. Evidently, $\Omega(-k_x,-k_y)=-\Omega(k_x,k_y)$ as a manifestation of time-reversal symmetry.   Lower Panel: (c) and (d) are the contour plots of the derivatives of $\Omega$ with respect to  $k_y$ and $k_x$, respectively. Since $\partial_y\Omega(k_x,k_y)=-\partial_y\Omega (k_x,-k_y)$, the integral over $k_y$ at zero temperature is found to be zero, leading to zero Berry curvature dipole along $y$ direction. In contrast, $\partial_x\Omega(k_x,k_y)=\partial_x\Omega (-k_x,k_y)$ which in turn leads to finite Berry dipole. The white spaces in all plots represent very large values of Berry curvatures (peaks or dips in 3D plots).}
	\label{fig:berryC}
\end{figure}

{\em Berry curvature dipole in 2D Dirac semimetals:} 
The simple form of Eq.~\ref{eq:SemiDirac} in terms of fictitious magnetic field $\mathbf{d}(\mathbf{k})$ allows us to write Berry curvature as 
%The Berry curvature, which describes the bending of the parameter space, arising from the geometrical structure of quantum eigen states, acts as a fictitious magnetic field  in the momentum space. With this fictitious field, we can define the Berry curvature $\Omega({\bf k} )$ as 
\begin{align}
\Omega^{a}({\bf k})=\epsilon_{abc}\frac{1}{2 |\mathbf{d}(\mathbf{k})|^{3}}\mathbf{d}(\mathbf{k})\cdot\Big{(} \frac{\partial \mathbf{d}(\mathbf{k})}{\partial k_b} \times \frac{\partial \mathbf{d}(\mathbf{k})}{\partial k_c} \Big{)},
\label{bc}
\end{align}
where $\epsilon_{abc}$ is the usual Levi-Civita and $(abc)\in(xyz)$. With this, the Berry curvature dipole as characterized by the first-order moment of the Berry curvature over the occupied states is defined as
\begin{align}
D_{ab}=\int_{k}f_{o}(\partial_{a}\Omega_{b}),
\label{bdip}
\end{align}
where $\int_k=\int d^dk/(2\pi)^d$ and $f_{o}$ is the Fermi-Dirac distribution function. 
%The nonlinear Hall effect has a quantum nature because of its connection with the Berry curvature dipole.
In three-dimension (3D), the Berry curvature $\bm{\Omega_k}$ is a pseudovector and consequently, the BCD ($D_{ab}$) becomes a pseudotensor. On the other hand, in the case of a two-dimensional system, the only nonzero component of the $\bm{\Omega}$ is $\Omega^{z}({\bf k})$ $\left(a=z\right)$, indicating the fact that the Berry curvature behaves as a pseudoscalar. Thus in 2D, the pseudotensorial quantity $D_{ab}$ is reduced to a pseudovector quantity ($D_{a}$) confined in the 2D plane with only two independent ($x$ and $y$) components.  We point out that for finite Berry curvature dipole to exist, the system must have at most one mirror symmetry left where the single mirror symmetry forces BCD to be orthogonal to the mirror line (i.e. a mirror plane that is perpendicular to the 2D system).

%Compared with the BCD \cite{Fu2015_NLAHE} defined as $D_{ab}= \int [d \bm{k}] f_0 \partial_a \bm{\Omega}_b$, the NLAN coefficient $\bm{\Lambda}^T_a$is accompanied with an extra term $(E_{\bm{k}}-\mu)^2/T^2$ which is an even function with respect to $\bm{k}$.
%In two dimension, the Berry curvature has only one non-zero out-of plane component which can be viewed as ${\bf \Omega(\bf k)}=\Omega ({\bf k})\hat z $, where $\Omega({\bf k} )$ is defined as
%
%\begin{align}
%\Omega({\bf k})=\frac{1}{2 |\mathbf{d}(\mathbf{k})|^{3}}\mathbf{d}(\mathbf{k})\cdot\Big{(} \frac{\partial \mathbf{d}(\mathbf{k})}{\partial k_x} \times \frac{\partial \mathbf{d}(\mathbf{k})}{\partial k_y} \Big{)}.
%\label{bc}
%\end{align}
It is easy to see from Eq.~(\ref{bc}) that $\Omega({\bf k})=0$ for both the bands in Eq.~(\ref{eq:SemiDirac}) since $d_z=0$. This is attributed to the fact that Eq.~(\ref{eq:SemiDirac}) preserves all relevant symmetries protecting the gapless point. To generate finite Berry curvature, we introduce a perturbation $\delta h=m_0\,\sigma_z$ to ${\mathcal H}$, which breaks both $\mathcal{P}$ and $M_y$ symmetry but preserves $M_x$ and time-reversal symmetry. This gives
\begin{align}
\Omega({\bf k})=\frac{\beta\,k_x}{E_{\bf k}^3},
\label{eq:berryC}
\end{align}
where $E_{\bf k}=\sqrt{|\mathbf{d}(\mathbf{k})|}$ and $\beta=2\,\alpha\,v\,m_0$. Since $\Omega({\bf k})=-\Omega({\bf -k})$, the integral of the Berry curvature over the entire Brillouin zone, namely the Chern number turns out to be zero, hence we obtain zero linear anomalous Hall conductivity. It is also apparent from the right top panel of Fig.~(\ref{fig:berryC}) . For comparison, we have also shown Berry curvature for a isotropic Dirac Hamiltonian ($\sigma\cdot {\bf k}$) in the left top panel of  Fig.~(\ref{fig:berryC}). Since $\Omega({\bf k})=\Omega(-{\bf k})$, the integral of $\Omega$ over the entire Brillouin zone is expected to be finite. However, whether the total Chern number finite or zero depends on the contribution coming from all inequivalent Dirac nodes of the respective lattice model. For example, in graphene, the perturbation $\delta h$, namely "Semenoff mass" arising from the staggered on-site potential between two sublattices gives rise to insulting phase as the Chern numbers for the two inequivalent {\it gapped} Dirac nodes are equal but opposite in sign. We note that such mass term in deformed graphene and other possible candidate materials can be induced by light with different polarization. 
\begin{table}[t]
	\begin{tabular}{ |c|c|c|c|c|c| }
		\hline
		Material & $m/m_e$ & $m_0$ (eV)&$\mu$ (eV) &$D_x({\rm nm})$\\ 
		\hline
		(TiO$_2)_5$/(VO$_2)_3$			&$13.6$ 	& $0.2$	&$0.25$ 	&$0.27$ 		 \\  
		$\alpha$-(BEDT-TTF)$_2$I$_3$	&$3.1$ 	& $0.1$	&$0.15$ 	&$0.86$ 		\\
		Photonic crystals			 	&$1.2\times 10^{-3}$ &$1.0$	&$1.5$	&$13.0$		\\ 
		\hline
	\end{tabular}
	\caption{\label{tab:params}Microscopic parameters for semi-Dirac materials, with representative gaps $(m_0)$, chemical potentials, effective masses ($m$) and corresponding Berry curvature dipoles\cite{pardo2009pickett,Kaplunov,wu2014}. Here  $m_e$ represents the free electron mass. Evidently, the $D_x$ increases with the decrease in effective band masses as explained in the main text.}
\end{table}

%\textcolor{red}{\bf Please discuss about the mirror symmetry of this Hamiltonian.}
At zero temperature, the BCD can be computed using Eq.~\ref{bdip}, where the momentum integral is restricted to the region $E_k<\mu$. For the present model, the Fermi surface topology adds complexities in finding the analytical results using Eq.~\ref{bdip}. Thus we rewrite Eq.~(\ref{bdip}) as $D_i=-\int_k v_i\,\Omega^z\,f_0'$, where $v_i=\nabla_{k_i}E_{\bf k}$ with $i\in (x,y)$ and $'$ denotes the derivative with respect to the energy. With this, we obtain     

%In two dimension (2D) Berry curvature dipole behaves as a psuedovector and lies in 2D plane. It is defined as\cite{inti}
%\begin{align}
%D_{i}=\int_{k}f_{o}(\partial_{i}\Omega_{z}),
%\label{bdip}
%\end{align}  
%where $\int_k=\int d^2k/(2\pi)^2$, $f_{o}$ is the Fermi function and $i$ represents the direction in the plane. Eq.~(\ref{bdip}) can further be recast as $D_i=\int_k v_i\, \Omega_z f_0'$, where $v_i$ is the velocity along $i$ direction and $'$ denotes derivative with respect to energy. With this, the Berry curvature dipole at low temperature is obtained to be
\begin{align}
&D_x=2\,m_0\,\sqrt{\alpha} \frac{\sqrt{\mu^2-m_0^2}}{\mu^3}\mathcal{I}(\mu,\delta_0),\nonumber\\
&D_y=0,\label{eq:analytical}
\end{align}
where $\mathcal{I}(\mu,\delta_0)=\int_{0}^{2\pi}d\theta\,\left(F^{+}-F^{-}\Theta ({F^{-}}^2)\right)|\cos\theta|$ with $F^{s}= \left(s\sqrt{(\mu^2-m_0^2)}|\cos\theta|+\delta_0\right)^{1/2}$, $\Theta(x)$ is the usual Heaviside function. For $\delta_0=0$, $\mathcal{I}(\mu,\delta_0)$ can be further simplified and Eq.~\ref{eq:analytical} reads off
\begin{align} 
D_x=2\,m_0\,\sqrt{\alpha}\,\mathcal{I}_0 \frac{(\mu^2-m_0^2)^{3/4}}{\mu^3},
\end{align}
where ${\mathcal I}_0\simeq3.5$.
%\mathcal{I}(\mu,0)=\frac{2(1+i)}{3}&(\mu^2-m^2)^{1/4}\nonumber\\\times [(i+1)\,  &\mathcal{EK}(2)+i\sqrt{2}\,\mathcal{EK}(1/2)],
%\end{align}
%where $\mathcal{EK}$ is the EllipticK function as defined in WOLFRAM MATHEMATICA. 
To obtain Eq.~(\ref{eq:analytical}), we have used the parametrization $ k_x= \text{sign}[\cos\theta] \left( \frac{r \,|\cos\theta|+\delta_0}{\alpha} \right )^{1/2}$ and $ k_y= \frac{ r \, \sin\theta}{v}$\cite{mandal_2020}. Note that the Berry curvature dipole along $x$ survives due to the surviving mirror symmetry $M_x$ while $D_y=0$ as a consequence of broken $M_y$. Moreover, $D_x$ vanishes if $\mu$ lies inside the bulk gap i.e. $\mu\le m_0$. Notice that $D_x$ is independent of Dirac velocity $v$, which is in sharp contrast to the BCD for typical Dirac dispersion with a tilt or warping terms\cite{Fu_2015,Carmine_2019}. Also, it is indeed apparent that for materials with small band gap and small effective mass, the BCD is expected to be very large. In Table~\ref{tab:params}, we have presented typical mass parameters with representative gaps and corresponding Berry curvature dipoles for proposed semi-Dirac materials. Evidently, the BCD in a semi-Dirac material is comparable or larger than the BCD in materials with tilting  such as SnTe ($D\sim3\, nm$), TMDCs ($D\sim 10^{-2}\, nm$ )\cite{Fu_2015} or materials with warping such as graphene ($D\sim 10^{-3}\,nm$)\cite{Carmine_2019}.
\begin{figure}
	\includegraphics*[width=0.99\linewidth]{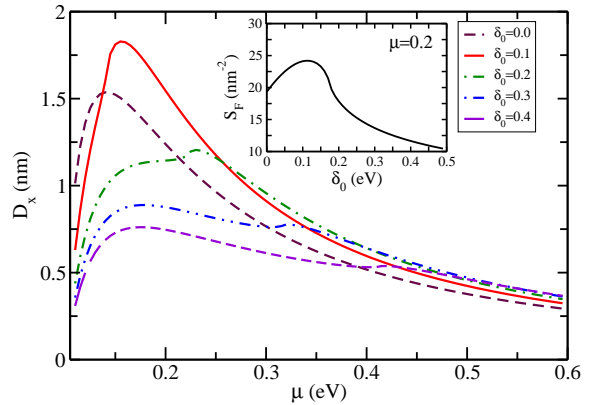}
	\caption{Berry curvature dipole (BCD) as a function of chemical potential for different values of $\delta_0$. Clearly for low doping, the BCD increases as the Dirac nodes move close to each other ($\delta_0\rightarrow 0$). This is attributed to the enhancement of the area of the Fermi surface ($S_F$) as evident from Fig.~(\ref{fig:band}) and the inset of Fig.~\ref{fig:berrydipole}. However, the Berry dipole suddenly reduces as the two Dirac nodes merge ($\delta_0=0.0$) into a single one, corroborating the nonmonotonic nature of the $S_F$ (see inset). For high doping, the  $S_F$ does not change substantially with $\delta_0$, leading to similar  $D_x$. Here, we have used $m_0=0.1$ eV and $m/m_e=13.6$ for 	(TiO$_2)_5$/(VO$_2)_3$.} % The black dot curve dentoes the analytical solution at $\delta_0=0$. }
	\label{fig:berrydipole}
\end{figure}

Figure \ref{fig:berrydipole} demonstrates the Berry curvature dipole with the variation of $\delta_0$. For low doping, $D_x$ increases as $\delta_0$ decreases. However, at $\delta_0=0$, the $D_x$ suddenly reduces compared to the $D_x$ at $\delta_0>0$ as apparent from the Fig.~(\ref{fig:berrydipole}). The is because the area of the Fermi surface, $S_F$ changes {\it nonmonotonically} with $\delta_0$ as illustrated in the inset of Fig.~(\ref{fig:berrydipole}). This behavior can further be understood from the Fig.~(\ref{fig:band}). For finite $\mu$ in the conduction band, we have a single Fermi surface at $\delta_0=0$ since the band has only one minima (refers to semi-Dirac node). As we move away from $\delta_0=0$, the single minima splits into two  minimas (refers to Dirac nodes) and a saddle point appears between them. However, the single Fermi surface retains as long as $\delta_0<\sqrt{\mu^2-m_0^2}$. Accordingly, the area increases due to the additional curvature arising from the spliting of the single minima. As we further increase $\delta_0$, the saddle point crosses $\mu$ at $\delta_0=\sqrt{\mu^2-m_0^2}$, leading to two Fermi surfaces. Consequently, the area starts to decrease. If we increase $\delta_0$ even more,  the bands near the Dirac nodes become narrower, which results in the reduction of $S_F$. For high doping, the Fermi surface topology almost remains same irrespective of the values of $\delta_0$, leading to similar $D_x$. Note that this feature of BCD differs from the case for fixed $\delta_0$ but varying chemical potential where $S_F$ monotonically increases with $\mu$ (not shown here). We also note that the BCD is found to have a little kink at $\mu=\sqrt{\delta_0^2+m_0^2}$ as a manifestation of topological Lifshitz transition.  %In contrast, Berry dipole along $y$ direction vanishes indeed due to the typical Dirac dispersion without any tilting effect. 

\begin{figure}
	\includegraphics*[width=0.99\linewidth]{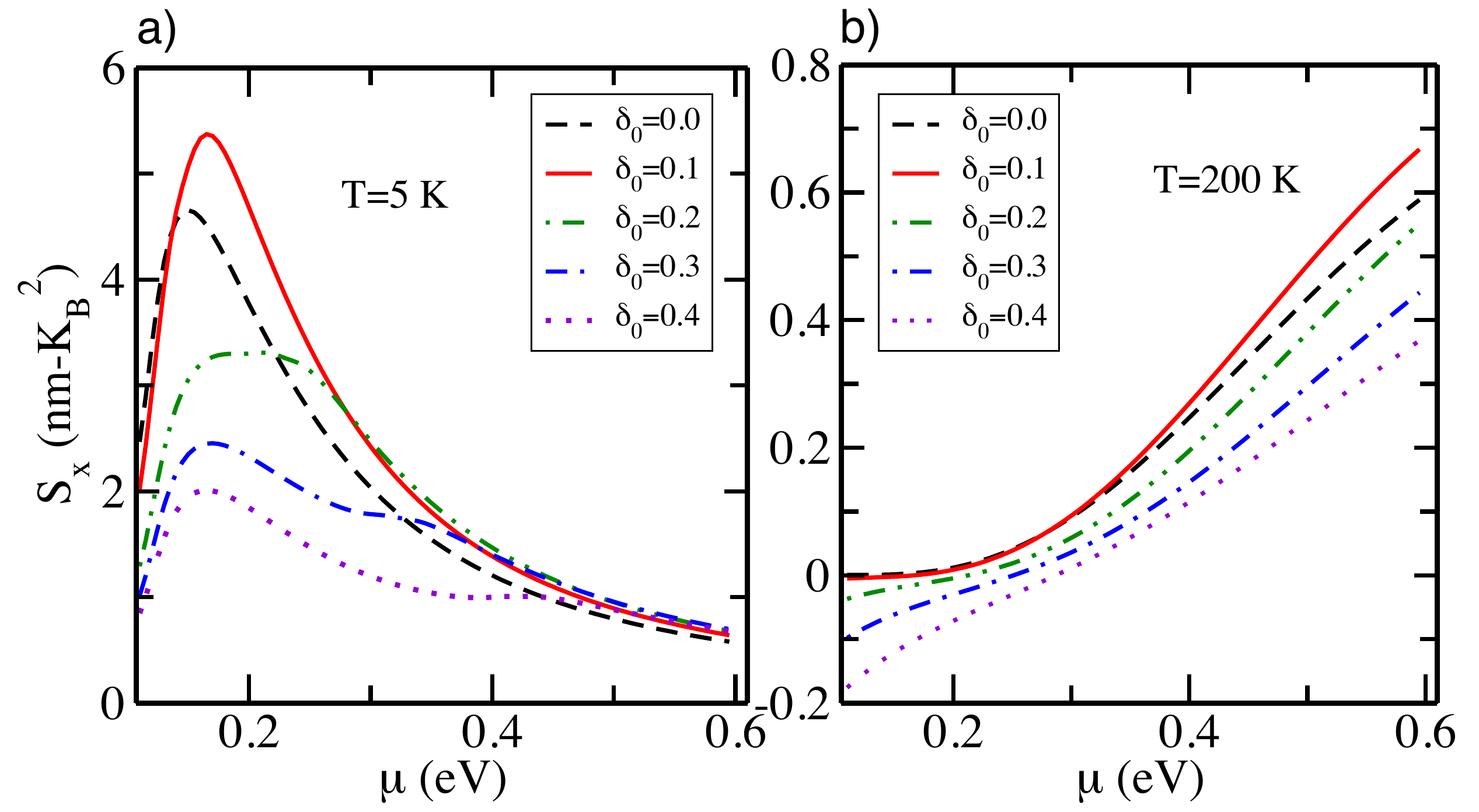}
	\caption{a) Nonlinear Nernst coefficient ($S_x$) as a function of chemical potential ($\mu$) for different values of $\delta_0$. It is apparent that $S_x$ mimics the $D_x$ even in the presence of additional term  $(E_k-\mu)^2$ at low temperature (T=5 K). b) The same plot for T=200 K. All parameters are used same as Fig.~\ref{fig:berrydipole}.}
	\label{fig:nonlinearnernst}
\end{figure}

{\em Non-linear photo current:}
We next move to the Berry curvature contribution to the dc photocurrent. 
In the presence of an external electric field (${\mathcal E}e^{i\omega t}$) with ${\mathcal E}\in{\mathbb C}$, the non-linear current is produced due to the Berry phase of Bloch electrons. In particular, the anomalous velocity of the Bloch electrons gives rise to a net current:
\begin{align}
{\bf j}=e\int_k \left[\mathbf {\dot{k}}\times {\mathbf \Omega(k)}\right]f_{neq}({\bf k}),
\label{eq:current} 
\end{align}
where $f_{neq}({\bf k})$ is the non-equilibrium electron distribution function and $\mathbf {\dot{k}}={\rm Re}\,(\frac{e}{\hbar}{\mathcal E}e^{i\omega t})$. In the relaxation time approximation with energy-independent scattering time $\tau$, the non-equilibrium electron distribution is obtained to be\cite{Moore_2015} 
\begin{align}
f_{neq}({\bf k})=2\,\tau\,e\,f'_0\frac{\left({\rm Re}\,{\mathcal E}-\omega\,\tau {\rm Im}\, \mathcal E\right)\cdot v}{1+\omega^2\,\tau^2}
\label{eq:nonequidist}
\end{align} 
Assuming low temperature and low frequency, Eq.~(\ref{eq:current}) together with Eq.~(\ref{eq:nonequidist}) leads to 
\begin{align}
%j=\chi_{z}\left[\frac{1}{\tau}\,({\mathcal E}_y {\mathcal E}_x^{\ast}+{\mathcal E}_x {\mathcal E}_y^{\ast})\,\hat x+i\, \omega\,({\mathcal E}_x {\mathcal E}_y^{\ast}-{\mathcal E}_y {\mathcal E}_x^{\ast})\,\hat x\right.\nonumber\\
%\left.-\frac{2}{\tau}|{\mathcal E}_x|^2\, \hat y\right],
{\bf j}=\frac{\chi}{1+\omega^2\,\tau^2}\left[\frac{1}{\tau}\,[{\mathcal E}_y {\mathcal E}_x^{\ast}]_+\hat x+i\,\omega\,[{\mathcal E}_y {\mathcal E}_x^{\ast}]_-\hat x-\frac{2}{\tau}|{\mathcal E}_x|^2\, \hat y\right],
%\left.-\frac{2}{\tau}|{\mathcal E}_x|^2\, \hat y\right],
\label{eq:curr}
\end{align}
where $[AB^{\ast}]_{\pm}=A B^{\ast}\pm A^{\ast}B$ and $\chi$ is obtained to be
\begin{align}
\chi=\frac{e^3}{\hbar}\int \frac {d^2 k}{2\pi}\, v_x\,\Omega^z ({\bf k})\,\delta (\mu-E_{\bf k})
=\frac{e^3}{\hbar}\,D_x.
\end{align}
The first two terms in Eq.~(\ref{eq:curr}) are known  as linear (LPGE) and circular (CPGE) photogalvanic currents, respectively. The last term in Eq.~(\ref{eq:curr}) denotes typical photovoltaic effect. It is clear that the CPGE changes sign with the helicity of the light wave and it is maximum for circularly polarized light. In contrast,  the LPGE is dominant and maximum for linearly polarized light. We further see that the non-linear conductivity is nothing but the Berry curvature dipole, corroborating the relation between photocurrent and Berry dipole as expected\cite{Fu_2015}. For a laser power of 1 Watt, $\omega\,\tau=1$, $m=13.6 m_e$ ((TiO$_2$)$_5$/(VO$_2$)$_3$),  the current density is found to be roughly of the order of $300 nA/mm$, which can be easily measured in standard experiments.  

To this end, we note that the anomalous velocity associated with the Berry curvature gives rise to helicity-dependent photocurrent if $\Omega({\bf k})\propto {\bf k}$ as pointed out in Ref.~\cite{Moore_2015}. It has been shown that while the Berry phase contribution vanishes in the bulk of a semiconductor quantum well, the surface confinement leads to the helicity-dependent photocurrent due to the particular nature of the Berry curvature. Interesting, our present model naturally gives rise to the helicity-dependent photocurrent without any external perturbation as $\Omega=\beta\, k_x/E_{\bf k}^3$ (cf. Eq.~\ref{eq:berryC}), in contrast to the typical Dirac dispersion.

{\em Non-linear Anomalous Nernst Effect:}
Next, we turn to non-linear anomalous Nernst effect in this system. Using the semiclassical Boltzmann theory within the relaxation time approximation, the non-linear anomalous Nernst coefficient at temperature $(T)$ can be defined as~\cite{Su_2019,Nandy1_2019}
\begin{align}
S_i=\int_k \frac{(E_{\bf k}-\mu)^2}{T^2}v_i\,\Omega({\bf k}) f'_0.
\label{eq:nernst}
\end{align}
The NLANE (second-order
response function to the applied temperature gradient), which refers to the nonlinear current, flowing perpendicular to the temperature gradient even in the absence of a magnetic field, is induced by the BCD and therefore, only $x$-component of it i.e., $S_x$ is finite in this system. 
%Eq.~(\ref{eq:nernst}) can further be simplified as 
%At low temperature and for $\mu\ll k_B T$, we obtain leading order temperature dependent term
%\begin{align}
%S_x=\frac{1}{T^2}\int_{0}^{2\pi}d\theta\left(\int_0^{\mu} dr-\int_{\delta_c}^{0}dr\right) \, h(r,\theta) f'_0(r),
%\label{eq:nernst1}
%\end{align}
%where $h(r,\theta)=r^2\,(\sqrt{r^2+m^2}-\mu)^2\,(r |\cos\theta|+\delta_0)^{1/2}/(r^2+m^2)^{3}|\cos\theta|$ and $\delta_c=-\delta_0/|\cos\theta|$. 
%At $\delta_0=0$, the integral in Eq.~(\ref{eq:nernst}) reduces to 
%\begin{align}
%S_x=\frac{1}{T^2}[(\mu^2-m^2/2)\sinh^{-1}(\mu/m)+\frac{\mu}{2}(-\mu m^2\nonumber\\+\frac{m^2-\mu^2}{\sqrt{m^2+\mu^2}}8\tan^{-1} (\mu/m))]
%\end{align} 
In Fig.~(\ref{fig:nonlinearnernst}), we show the behavior of Nernst coefficient at two different temperatures $T=5$ K and $T=200$ K. The qualitative feature turns to be same as BCD at low temperature even in the presence of the additional term $(E_K-\mu)^2$ in Eq.~(\ref{eq:nernst}). However, at high temperature it deviates substantially from the low temperature behavior as evident from Fig.~(\ref{fig:nonlinearnernst}). Notice that in both cases, the non-monotonicity of the area of the Fermi surface as discussed before is reflected.

{\em Conclusion:}
To conclude, we have identified a simple platform to observe substantial non-linear transport phenomena arising from the Berry curvature dipole (BCD). Specifically, we have shown that the sizable BCD can be obtained in a low-energy Dirac Hamiltonian with two Dirac nodes close to each other or their merging at a single node, namely the semi-Dirac node. Remarkably, the BCD for the present model is found to be independent of the Dirac velocity and predominantly depends on the inverse of effective mass of the Dirac quasiparticles as compared to the typical isotropic Dirac systems with well separated Dirac nodes.  Indeed, this is one of the interesting findings of this study. For typical Dirac materials, $v$ is of the order of $10^5 m/s$ and does not vary significantly from one material to another. However, the effective masses may vary significantly as evident from the Table I of possible semi-Dirac systems. 
%Thus having only effective mass-dependent BCD helps us to identify materials with significantly large Berry curvature dipole than the only velocity dependent BCD. 
This fact may guide us to identify materials with low effective masses responsible for a sizable BCD than the only velocity dependent BCD. For example,
%We point out that this fact can be used to  or 
the candidate semi-Dirac materials have very low effective masses  (Tabel.~\ref{tab:params}), hence they are potential platforms to observe BCD-induced large non-linear transport phenomena. We further show that the present model naturally host the helicity-dependent photocurrent due to the linear momentum-dependent Berry curvature in contrast to the isotropic Dirac dispersion. Finally, we present non-linear Nernst effect, arising solely due to the Berry curvature effect. Since the NLAHE can transform ac electric fields into dc currents, a process known as rectification\cite{Fu_2020}, the proposed Dirac materials may have great potential applications for next-generation wireless and energy-harvesting devices.      

{\em Acknowledgement:} KS is thankful to A. Jaiswal for useful discussion.

\bibliography{berrydipole}{}

\begin{thebibliography}{40}
\bibitem [{not()}]{note_1}%
  \BibitemOpen
  \href@noop {} {}\bibinfo {note} {The Fermi surface topologies represent here
  an equation of a contour satisfying $E(k_x,k_y)=\mu$, where $E(k_x,k_y)$ is
  the energy spectrum of the conduction band.}\BibitemShut {Stop}%
\end{thebibliography}

%--------------------------------------------------------------------------
\end{document}